\documentclass[a4paper, 10pt, conference]{ieeeconf}      

\IEEEoverridecommandlockouts                              
\usepackage{graphics} 
\usepackage{amsmath} 
\usepackage{amssymb}  
\usepackage{pgfplots}
\pgfplotsset{compat=newest} 
\usepackage{graphicx,caption,subcaption}
\usepackage{wrapfig}

\usepackage{amsthm}
\newtheorem{theorem}{Theorem}
\newtheorem{definition}{Definition}
\newtheorem{example}{Example}

\newtheorem{lemma}{Lemma}

\usepackage{tikz,tikzscale}
\usetikzlibrary{matrix, fit, positioning}
\usetikzlibrary{backgrounds}

\usepackage{sidecap}

\newif\ifcomment
\commenttrue
\usepackage{color}

\newcommand{\field}{\mathbb{F}}
\newcommand{\ceil}[1]{\left\lceil{#1}\right\rceil}
\newcommand{\floor}[1]{\left\lfloor{#1}\right\rfloor}
\newcommand{\code}{\mathcal{C}}
\DeclareMathOperator{\wt}{wt}
\DeclareMathOperator{\vd}{vd}
\DeclareMathOperator{\vdb}{VDB}
\newcommand{\defeq}{\mathrel{\mathop=^{\textrm{def}}}}
\newcommand{\defref}[1]{Definition~\ref{#1}}
\newcommand{\thref}[1]{Theorem~\ref{#1}}

\newcommand{\figref}[1]{Figure~\ref{#1}}
\newcommand{\egref}[1]{Example~\ref{#1}}
\newcommand{\secref}[1]{Section~\ref{#1}}

\title{\LARGE \bf
Locality in Crisscross Error Correction}

\author{Hedongliang Liu, Lukas Holzbaur and Antonia Wachter-Zeh
\thanks{H. Liu, L. Holzbaur and A. Wachter-Zeh are with the Insitute for Communications Engineering, Technical University of Munich, Germany.
{\tt\small Emails: \{lia.liu, lukas.holzbaur, antonia.wachter-zeh\}@tum.de}}%
\thanks{L. Holzbaur's and A. Wachter-Zeh's work was supported by the Technical University of Munich-Institute for Advanced Study, funded by the German Excellence Initiative and European Union Seventh Framework Programme under Grant Agreement No. 291763 and the German Research Foundation (Deutsche Forschungsgemeinschaft, DFG) unter Grant No. WA3907/1-1.}
}%

\begin{document}


\maketitle
\thispagestyle{empty}
\pagestyle{plain}
\pagenumbering{gobble}

\begin{abstract}
The {cover metric} is suitable for describing the resilience against correlated errors in arrays, in particular crisscross errors, which makes it interesting for applications such as distributed data storage (DDS). In this work, we consider codes designed for the cover metric that have {locality}, that means lost symbols can be recovered by using only a few other (local) symbols.
We derive and prove a Singleton-like bound on the minimum {cover distance} of {cover-metric codes} with locality and propose a bound-achieving construction. 
Further, we explore the performance of our construction in comparison to a known construction based on rank-metric codes.
\end{abstract}

\section{Introduction}
Distributed and cloud storage systems have reached such a massive scale that recovery from server failures is now part of regular opertation of a system rather than a rare exception. 
Although today's storage systems are resilient to several concurrent node failures, by far the most common scenario are failures of a single node. Hence, a storage system should be designed to efficiently repair these likely cases. The repair efficiency of a single node failure in a storage system can be quantified under different metrics, where the relevance is determined by the specific storage systems and applications. A large body of existing work has considered the repair problem under three criteria: i) repair-bandwidth~\cite{RB1}
, i.e., the number of bits communicated in the network, ii) the number of bits read~\cite{numread1} and iii) repair locality~\cite{locality1,AFamily}, i.e., the number of nodes that participate in the repair process. 

Most of the existing work has considered block codes in the Hamming metric. In some applications, however, failures can be bursts of errors which are correlated rather than independent. For instance, breakdown or simultaneous upgrades of several servers or damage of a rack switch or a power supply in distributed data storage systems can be deemed to be highly-correlated failure patterns~\cite{Ford10,Gill11}. Similar failure patterns can be found in dynamic random-access memories (DRAMs). A typical DRAM chip contains several internal \emph{banks}, each of which is logically organized into \emph{rows} and \emph{columns}. Each row/column address pair identifies a \emph{word} composed of several bits. It has been shown by several studies that DRAMs suffer from non-negligible percentage of single-row errors, single-column errors and single-bank errors~\cite{Srid15,DRAM09}. 
To cope with these particular failure patterns in applications, the following error correction problem needs to be considered: the data to be stored is divided into $n\times n$ arrays, with the possibility that some symbols are recorded erroneously. The error positions are highly correlated due to the structure of the memories. We can then get an abstraction of the error patterns such that all corrupted symbols are confined to several rows or columns (or both). Such an error model is referred as \emph{crisscross errors}~\cite{Roth-Crisscross,gabi85opt}. 

The characteristic of error models can be measured in different metrics.
A suitable metric with respect to crisscross errors is the \emph{cover metric}~\cite{Roth-Crisscross,Wachter-Zeh-Crisscross}, also called \emph{2D-burst metric} in~\cite{Gab2D12}. In the cover metric, the distance between two matrices of equal size is the number of rows/columns in which they differ. Array codes with a high distance in this metric based on \emph{maximum distance separable} (MDS) codes were constructed by Roth in~\cite{Roth-Crisscross}. However, similar to the Hamming metric, using MDS codes results in high repair costs of this construction for a single node failure.
In~\cite{S.Kadhe}, a family of codes with locality in the \emph{rank metric} was constructed, \emph{i.e.}, the distance between two matrices of equal size is the rank of the difference matrix~\cite{Gabi85}. As the cover and the rank of a matrix are related, this construction also has a high distance and some locality properties in the cover metric. However, the construction from~\cite{S.Kadhe} only guarantees small repair sets for columns, not for rows.  

In this work, we focus on codes with locality in the cover metric. \secref{sec:preliminaries} gives some preliminaries on the {cover metric} and locality. In \secref{sec:codes}, we give a definition of locality in the cover metric. Further, we prove a Singleton-like bound on the minimum cover distance of codes with \emph{cover-locality} and give a construction that achieves the bound with equality. Finally in \secref{sec:prob}, we analyze the probability of locally decoding crisscross errors/erasures of certain cover weight with our construction and compare it to the rank-metric construction from \cite{S.Kadhe}.

\section{Preliminaries}\label{sec:preliminaries}
\subsection{Notation}
Let $q$ be a power of a prime and denote by $\mathbb{F}_q$ the finite field of order $q$.
We write $[n]$ for the set of integers $\{1,2,\hdots,n\}$. \par
We denote a linear code over $\mathbb{F}_q$ of length $n$, dimension~$k$ and minimum distance $d$ by $[n,k,d]_q$, the weight of an element $\mathbf{c}$ in the respective metric by $\wt(\mathbf{c})$ and the minimum distance of a code $\mathcal{C}$ by $d(\mathcal{C})$. If we consider a specific metric, it is indicated by $H$, $C$ or $R$ for the Hamming, cover, and rank metric respectively (\emph{e.g.}, we write $d_H(\code)$ for the Hamming distance of a $[n,k,d]_q^H$ code $\code$). 
Let $S$ be a set of coordinates/tuples, then $\mathcal{C}|_{{S}}$ denotes the code obtained by restricting the code $\mathcal{C}$ to the coordinates/tuples of $S$.

\subsection{Cover Metric Codes}
In the cover metric, rows and columns of a matrix are treated indifferently. We use the term \emph{lines} to address the rows or columns of a matrix. In other words, we say that an $m\times n$ matrix has $m+n$ lines rather than $m$ rows and $n$ columns.\par
Let $\mathbf{E} \in \field_q^{m\times n}$ and $e_{ij}$ be the entry of $\mathbf{E}$ in the $i$-th row and $j$-th column. A \emph{cover} of $\mathbf{E}$ is any set $\mathcal{X}\subseteq[m+n]$ such that ${e}_{ij}\neq 0 \Rightarrow (i\in\mathcal{X} \,\mathrm{or}\, j+m\in\mathcal{X}) \; \forall \; i\in[m],j\in[n]$. Let $C(\mathbf{E})$ be the set of all possible covers of $\mathbf{E}$. The \emph{cover-weight} of $\mathbf{E}$ is defined as $$\wt_C(\mathbf{E})=\min_{\mathcal{X}\in C(\mathbf{E})} |\mathcal{X}|.$$
Note that the minimum-size cover of a matrix may not be unique.
The \emph{cover distance} between two matrices $\mathbf{A}, \mathbf{B}\in\mathbb{F}_q^{m\times n}$ is the minimum cover weight of the difference between the two matrices: 
\begin{equation*}
d_C(\mathbf{A,B})\defeq \wt_C(\mathbf{A}-\mathbf{B}).
\end{equation*}

A linear \emph{cover-metric code} $\mathbb{C}$, denoted by $[n\times n, k, d]_q^C$, 
is a linear subspace of $\mathbb{F}_q^{n\times n}$ of dimension $k$ and \emph{minimum cover distance} $d$. The minimum cover distance is defined as
\begin{equation*}
d =\underset{\substack{\mathbf{A}\neq\mathbf{B}\\\mathbf{A},\mathbf{B}\in\mathbb{C}}}{\min}\wt_C(\mathbf{A}-\mathbf{B}) = \underset{\substack{\mathbf{A}\in\mathbb{C}\\\mathbf{A}\neq\mathbf{0}}}{\min}\wt_C(\mathbf{A}). 
\end{equation*}

\subsection{Locality}

The locality of a code is defined as the \emph{number of symbols} 
participating in the process of recovering a lost symbol. In particular, an $[n,k,d]_q$ code is said to have locality $r$ if every symbol is recoverable from at most $r$ other symbols.
\begin{definition}[Locality]\label{def:locality}
Given a metric $M$, an $[n,k,d]_q^M$ code $\mathcal{C}$ is said to have $(r,\rho)$ locality, if for each symbol ${{c}_i, i \in[n]}$ of a codeword $\mathbf{c} = ({c}_1, {c}_2,...,{c}_n)$, there exists a set $\Gamma(i)\subset[n]$ of indices such that
\begin{enumerate}
    \item $i\in\Gamma(i)$,
    \item $|\Gamma(i)|\leq r+\rho -1$, and
    \item $d_M(\mathcal{C}|_{\Gamma(i)}) \geq \rho$.
\end{enumerate}
\end{definition}
We denote an $[n,k,d]_q$ code with $(r,\rho)$ locality by $[n,k,d,r,\rho]_q$ and refer to it as a \emph{locally repairable code} (LRC).
For any $[n,k,d,r,\rho]_q^H$ code $\mathcal{C}$, the minimum Hamming-distance is bounded by (see \cite{Kamath-Codes}):
\begin{equation}\label{eq:LRC_bound}
    d_H(\mathcal{C})\leq n-k+1-\left(\ceil{\frac{k}{r}}-1\right) (\rho - 1).
\end{equation}

\section{Locality in Cover Metric}\label{sec:codes}
We are interested in cover-metric codes such that the local code associated with each local group is an array code with minimum cover distance guarantee. First we generalize the concept of locality to the cover metric. For ease of notation we only consider square matrices of size $n \times n$.
\begin{definition}[Cover-Locality]\label{def:coverlocality}
An $[n\times n,k,d]_q^C$ code $\mathbb{C}$ is said to have $(r,\rho)$ cover-locality if for any line $i$, $i\in[2n]$ of a codeword $\mathbf{C}\in\mathbb{C}$, there exists a set $\Gamma(i)\subset[n]$ of indices of \emph{non-intersecting} lines such that
  \begin{enumerate}
    \item $i\in\Gamma{(i)}$, 
    \item $|\Gamma(i)|\leq r+\rho-1$, and
    \item $d_{C}(\mathbb{C}|_{\Gamma(i)}) \geq \rho$.
  \end{enumerate}
\end{definition}
We refer to the restriction $\mathbb{C}|_{\Gamma(i)}$ of size $n\times |\Gamma(i)|$ or $|\Gamma(i)| \times n$ as a local code. Note that the most general definition of locality in cover metric is given by \defref{def:locality}, with lines as the symbols. However, it seems that this is not practical as the corruption of one line would imply the corruption of other lines in its recovery set, if the lines of a recovery set can intersect. We therefore require non-intersecting lines in the recovery sets, \emph{i.e.}, the recovery set of any row (column) consists only of rows (columns).
\subsection{Alphabet-free Bound on the Cover Distance}
We give an Singleton-like upper bound on the minimum cover distance of codes with cover-locality.
\begin{theorem}[Alphabet-free Bound on the Cover Distance]
\label{thm:CoverLocbound}
For any linear ${[n\times n,k,d,r,\rho]_q^C}$ array code~$\mathbb{C}$, the minimum cover distance is bounded by
\begin{equation}\label{eq:CoverLocbound}
    d_C(\mathbb{C})\leq n-\frac{k}{n}+1-\left(\ceil{\frac{k}{n{r}}}-1\right) (\rho - 1)
\end{equation}
\end{theorem}
\begin{proof}
Fixing a basis of $\mathbb{F}_{q^n}$ over $\mathbb{F}_q$ gives a bijective linear map $\phi(\cdot):\mathbb{F}^n_{q}\mapsto\mathbb{F}_{q^n}$, where $\phi(\mathbf{0}) \mapsto {0}$. This can be extended to a bijective map $\phi_{\textrm{ext}}(\cdot):\mathbb{F}^{n\times n}_q\mapsto\mathbb{F}^n_{q^n}$.
Then, for any matrix $\mathbf{C} \in \mathbb{F}^{n\times n}_q$, there is a corresponding vector $\mathbf{c}\in \mathbb{F}^n_{q^n}$ such that $\mathbf{c}=\phi_{\textrm{ext}}(\mathbf{C})$.
It holds that
\begin{equation}
\label{eq:cover-hamming}
\wt_C(\mathbf{C})\leq \wt_{\mathrm{column}}(\mathbf{C})=\wt_H(\mathbf{c}),
\end{equation}
where $\wt_{\mathrm{column}}(\mathbf{C})$ is defined as the number of non-zero columns of a matrix $\mathbf{C}$.
By this mapping every ${[n\times n,k,d]_q^C}$ array code $\mathbb{C}$ gives an ${[n,k'=\frac{k}{n},d']_{q^n}^H}$ block code $\mathcal{C}$. By the same argument, it holds that if $\mathbb{C}$ has $(r,\rho)$ cover-locality the corresponding code $\mathcal{C}$ has $({r},\rho)$ locality in Hamming metric. 
From \eqref{eq:cover-hamming} and the linearity of the codes it follows that 
\begin{equation*}
d_C(\mathbb{C})\leq d_H(\mathcal{C}).
\end{equation*}
An upper bound on the minimum Hamming-distance of $\mathcal{C}(n,k',d',r,\rho)^H_{q^n}$ is therefore also the upper bound on the minimum cover distance of $\mathbb{C}(n\times n,k,d,r,\rho)_q^C$. Hence, \eqref{eq:CoverLocbound} follows from \eqref{eq:LRC_bound} by substituting $k$ with $k'=\frac{k}{n}$.
\end{proof}
We say that a code with cover-locality is optimal if its minimum cover distance achieves \eqref{eq:CoverLocbound} with equality.

\subsection{Bound-Achieving Code Construction} 

We focus on codes with disjoint local codes of length $n_l=r+\rho-1$. Moreover, let $n_l \mid n$, $r \mid k$ and denote the number of local groups by $\mu\triangleq {n}/{n_l}$.

The idea of designing codes with cover-locality is inspired by the \emph{vector-diagonal $(\vd)$ construction}~\cite{Roth-Crisscross}, where the diagonals of the codewords of an array code are codewords of a linear block code. 
\begin{definition}[Vector-Diagonal Function]
For some integers $s$ and $n$ with $s\leq n$, given a set $\mathcal{T}$ of vectors $\mathbf{c}^{(i)} = ({c}^{(i)}_1, {c}^{(i)}_2, 
\dots, {c}^{(i)}_n)\in\mathbb{F}_q^n,\forall i\in[s]$, the $\vd$ function is defined as
\begin{equation*}
\begin{aligned}
\vd(\mathcal{T})\defeq
\left(
    \begin{smallmatrix}
    {c^{(1)}_{1}} &  &  & {c^{(s)}_{n-s+1}} & \hdots & {c^{(2)}_{n}}\\
    {c^{(2)}_{1}} & {c^{(1)}_{2}} &&& \ddots & \vdots \\
    \vdots &{c^{(2)}_{2}}& \ddots& & & {c^{(s)}_{n}}\\
    {c^{(s)}_{1}} & \vdots & \ddots & \ddots &  & &\\
    & \ddots & \vdots & \ddots & {c^{(1)}_{n-1}} & \\
    & & {c^{(s)}_{n-s}} & \hdots & {c^{(2)}_{n-1}} & {c^{(1)}_{n}}
    \end{smallmatrix}
\right).
\end{aligned}
\end{equation*}
\end{definition}
A straightforward approach for constructing codes with locality in the columns is to arrange codewords of a block LRC on the diagonals of each codeword array. However, the resulting code will not have cover-locality as defined in \defref{def:coverlocality} since the recovery sets for the rows do not fulfill the constraints.

To simplify the expression in the following code construction we first define a \emph{modified vector-diagonal} function.
\begin{definition}[Modified Vector-Diagonal Function]\label{def:vdb}
Let ${\mathcal{S}=\{\mathcal{S}_1,\mathcal{S}_2,...\}}$ be a partition of $[n]$; $\mathcal{T}$ be an ordered multiset of $n$ vectors $\mathbf{c}^{(s)}\in\mathbb{F}_q^n,\forall \; s \in[n]$; and ${\mathcal{T}_i = \{\mathbf{c}^{(s)}: s \in [(i-1)n_l+1,in_l] \}}$. Denote by $\mathcal{T}_{i,j}$ the multiset of vectors of $\mathcal{T}_i$ restricted to the coordinates of $\mathcal{S}_j$
\begin{equation*}
\mathcal{T}_{i,j} = \left\{\left.\mathbf{c}^{(s)}|_{\mathcal{S}_j} \; \right| \; \mathbf{c}^{(s)}\in \mathcal{T}_i \right\} .
\end{equation*}
Define the $\vdb$ function as
\begin{align*}
\vdb(\mathcal{T,S})
\defeq\vd(&[\vd(\mathcal{T}_{1,1}),\vd(\mathcal{T}_{1,2}),...,\vd(\mathcal{T}_{1,\mu})],\\
&[\vd(\mathcal{T}_{2,1}),\vd(\mathcal{T}_{2,2}),...,\vd(\mathcal{T}_{2,{\mu}})],...,\\
&[\vd(\mathcal{T}_{\mu,1}),\vd(\mathcal{T}_{\mu,2}),...,\vd(\mathcal{T}_{{\mu,\mu}})]).
\end{align*}
\end{definition}
See \egref{eg:vdbprocess} for an illustration. With the modified vector-diagonal function we can now give our code construction.
\begin{definition}[Codes with Cover-Locality]
\label{def:ArrayLRC}
Let $\code$ be a linear $[n, \frac{k}{n},d,{r}, \rho]^H_q$ block LRC and $\mathcal{S}=\{\mathcal{S}_1,...,\mathcal{S}_{\mu}\}$ be the corresponding partition into the local sets. Let $\mathcal{V}$ be the set of all ordered multisets $\{\mathbf{c}^{(1)},...,\mathbf{c}^{(n)}\}, \mathbf{c}^{(i)} \in \code$.
Define the code $\mathbb{C}$ as
\begin{equation*}
\mathbb{C} \defeq \{\vdb(\mathcal{T},\mathcal{S}): \mathcal{T} \in \mathcal{V} \},
\end{equation*}
where the function $\vdb$ is given in \defref{def:vdb}.
\end{definition}
We call $\mathcal{C}$ the \emph{constituent code} of $\mathbb{C}$.

\begin{theorem}\label{th:covLRC}
The code $\mathbb{C}$ as defined in \defref{def:ArrayLRC} is an optimal linear $[n\times n,k,d,r,\rho]_q^C$ cover-metric LRC.
\end{theorem}
\begin{proof}
The codeword size $n \times n$ and the dimension follow directly from the definition. The linearity of $\mathbb{C}$ follows from the linearity of its constituent code $\code$.
To show the minimum distance, consider a codeword $\mathbf{C}\in\mathbb{C}$, which is obtained from $n-1$ all-zero codewords and one non-zero $\mathbf{c}\in\mathcal{C}$ of the constituent code. The $\vdb$ function arranges all the symbols of $\mathbf{c}$ in distinct rows and columns of $\mathbf{C}$ and therefore $\wt_C(\mathbf{C})=\wt_H(\mathbf{c})$. Trivially increasing the number of non-zero constituent codewords can only increase the cover-weight and it follows that
\begin{equation*}
d_C(\mathbb{C}) = \underset{\substack{\mathbf{C}\in\mathbb{C}\\\mathbf{C}\neq\mathbf{0}}}{\min}\ \wt_C(\mathbf{C}) = \underset{\substack{\mathbf{c}\in\mathcal{C}\\\mathbf{c}\neq\mathbf{0}}}{\min}\ \wt_H(\mathbf{c}) = d_H(\mathcal{C}).
\end{equation*}
The locality follows by the same argument.
\end{proof}
\thref{th:covLRC} shows that if the constituent code achieves~\eqref{eq:LRC_bound}, the code as defined in \defref{def:ArrayLRC} is an optimal cover-metric LRC, \emph{i.e.}, achieves~\eqref{eq:CoverLocbound}.\par
\begin{figure}[h!]
	\begin{minipage}{0.5\linewidth}
    	\vspace{-2ex}
        \resizebox{\linewidth}{!}{
\begin{tikzpicture}
    \matrix[
        matrix of math nodes,
        row sep=.8ex,
        column sep=1ex,
        left delimiter=(,right delimiter=),
        nodes={text width=.75em, text height=.5ex, text depth=.5em, align=center}
        ] (m) 
        {
        \alpha_1   & \alpha_2   & \alpha_3   & \alpha_4   & \alpha_5   & \alpha_6 & \alpha_7 & \alpha_8 & \alpha_9\\
        \beta_1    & \beta_2    & \beta_3   & \beta_4   & \beta_5      & \beta_6 & \beta_7 & \beta_8 & \beta_9\\
        \gamma_1   & \gamma_2     & \gamma_3   & \gamma_4   & \gamma_5   & \gamma_6 & \gamma_7 & \gamma_8 & \gamma_9\\
        \epsilon_1 & \epsilon_2 & \epsilon_3 & \epsilon_4 & \epsilon_5 & \epsilon_6 & \epsilon_7 & \epsilon_8 & \epsilon_9\\
        \zeta_1    & \zeta_2    & \zeta_3     & \zeta_4    & \zeta_5   & \zeta_6 & \zeta_7 & \zeta_8 & \zeta_9\\
        \eta_1     & \eta_2   & \eta_3 & \eta_4   & \eta_5    & \eta_6 & \eta_7 & \eta_8 & \eta_9\\
        \pi_1      & \pi_2      & \pi_3   & \pi_4 & \pi_5     & \pi_6 & \pi_7 & \pi_8 & \pi_9\\
        \omega_1   & \omega_2   & \omega_3   & \omega_4    & \omega_5 & \omega_6 & \omega_7 & \omega_8 & \omega_9\\
        \kappa_1   & \kappa_2   & \kappa_3      & \kappa_4     & \kappa_5    & \kappa_6 & \kappa_7 & \kappa_8 & \kappa_9\\
        }; 
        \draw[dashed, thin, black] (m-1-4.north west) -- (m-9-4.south west);
        \draw[dashed, thin, black] (m-1-7.north west) -- (m-9-7.south west);
        \draw[dashed, thin, black] (m-3-1.south west) -- (m-3-9.south east);
        \draw[dashed, thin, black] (m-6-1.south west) -- (m-6-9.south east);

        \draw[orange,rounded corners] (-2.8,2.4) rectangle (-1,2.7);

        \draw[blue,rounded corners] (-2.8,1.8) rectangle (-1,2.1);

        \draw[red,rounded corners] (-2.8,1.2) rectangle (-1,1.5);

        \begin{scope}[on background layer]
            \node[fit=(m-1-7)(m-3-9), fill=green!35, rounded corners ] {};
            \node[fit=(m-1-4)(m-3-6), fill=green!25, rounded corners ] {};
            \node[fit=(m-1-1)(m-3-3), fill=green!15, rounded corners ] {};
            \node[fit=(m-4-7)(m-6-9), fill=yellow!35, rounded corners] {};
            \node[fit=(m-4-4)(m-6-6), fill=yellow!25, rounded corners] {};
            \node[fit=(m-4-1)(m-7-3), fill=yellow!15, rounded corners] {};
            \node[fit=(m-7-7)(m-9-9), fill=cyan!35, rounded corners  ] {};
            \node[fit=(m-7-4)(m-9-6), fill=cyan!25, rounded corners  ] {};
            \node[fit=(m-7-1)(m-9-3), fill=cyan!15, rounded corners  ] {};
        \end{scope} 
\end{tikzpicture}
}
	\end{minipage}%
	\hspace{0.5ex}
	\begin{minipage}{0.48\linewidth}
    	\vspace{-2ex}
		Step 1: Take an array with $n$ codewords of the constituent code as rows and partition it into $\mu\times\mu$ sub-arrays of size $n_l \times n_l$, where the partition in the columns is the partition of the constituent code into its local codes.
	\end{minipage}
	\begin{minipage}{0.5\linewidth}
		\vspace{1ex}
        \resizebox{\linewidth}{!}{
\begin{tikzpicture}
    \matrix[
        matrix of math nodes,
        row sep=.8ex,
        column sep=1ex,
        left delimiter=(,right delimiter=),
        nodes={text width=.75em, text height=.5ex, text depth=.5em, align=center}
        ] (m) 
        {
        \alpha_1 & \gamma_2 & \beta_3 & \alpha_4 & \gamma_5 & \beta_6 & \alpha_7 & \gamma_8 & \beta_9\\
        \beta_1 & \alpha_2 & \gamma_3 & \beta_4 & \alpha_5 & \gamma_6 & \beta_7 & \alpha_8 & \gamma_9\\
        \gamma_1 & \beta_2 & \alpha_3 & \gamma_4 & \beta_5 & \alpha_6 & \gamma_7 & \beta_8 & \beta_9\\
        \epsilon_1 & \eta_2 & \zeta_3 & \epsilon_4 & \eta_5 & \zeta_6 & \epsilon_7 & \eta_8 & \zeta_9\\
        \zeta_1 & \epsilon_2 & \eta_3 & \zeta_4 & \epsilon_5 & \eta_6 & \zeta_7 & \epsilon_8 & \eta_9\\
        \eta_1 & \zeta_2 & \epsilon_3 & \eta_4 & \zeta_5 & \epsilon_6 & \eta_7 & \zeta_8 & \epsilon_9\\
        \pi_1 & \kappa_2 & \omega_3 & \pi_4 & \kappa_5 & \omega_6 & \pi_7 & \kappa_8 & \omega_9\\
        \omega_1 & \pi_2 & \kappa_3 & \omega_4 & \pi_5 & \kappa_6 & \omega_7 & \pi_8 & \kappa_9\\
        \kappa_1 & \omega_2 & \pi_3 & \kappa_4 & \omega_5 & \pi_6 & \kappa_7 & \omega_8 & \pi_9\\
        }; 
        \draw[dashed, thin, black] (m-1-4.north west) -- (m-9-4.south west);
        \draw[dashed, thin, black] (m-1-7.north west) -- (m-9-7.south west);
        \draw[dashed, thin, black] (m-3-1.south west) -- (m-3-9.south east);
        \draw[dashed, thin, black] (m-6-1.south west) -- (m-6-9.south east);

        \draw[orange,rounded corners,rotate around={135:(-1.9,1.95)}] (-2.9,1.8) rectangle (-0.9,2.1);

        \draw[blue,rounded corners,rotate around={135:(-2.25,1.65)}] (-2.9,1.5) rectangle (-1.6,1.8);
        \draw[blue,rounded corners,rotate around={135:(-1.3,2.55)}] (-1.5,2.4) rectangle (-1.1,2.7);

        \draw[red,rounded corners,rotate around={135:(-2.6,1.35)}] (-2.8,1.2) rectangle (-2.4,1.5);
        \draw[red,rounded corners,rotate around={135:(-1.6,2.25)}] (-2.2,2.1) rectangle (-1,2.4);

        \begin{scope}[on background layer]
            \node[fit=(m-1-7)(m-3-9), fill=green!35, rounded corners ] {};
            \node[fit=(m-1-4)(m-3-6), fill=green!25, rounded corners ] {};
            \node[fit=(m-1-1)(m-3-3), fill=green!15, rounded corners ] {};
            \node[fit=(m-4-7)(m-6-9), fill=yellow!35, rounded corners] {};
            \node[fit=(m-4-4)(m-6-6), fill=yellow!25, rounded corners] {};
            \node[fit=(m-4-1)(m-7-3), fill=yellow!15, rounded corners] {};
            \node[fit=(m-7-7)(m-9-9), fill=cyan!35, rounded corners  ] {};
            \node[fit=(m-7-4)(m-9-6), fill=cyan!25, rounded corners  ] {};
            \node[fit=(m-7-1)(m-9-3), fill=cyan!15, rounded corners  ] {};
        \end{scope} 
\end{tikzpicture}
}
	\end{minipage}%
	\hspace{0.5ex}
	\begin{minipage}{0.48\linewidth}
		Step 2: Rearrange the entries of the subarrays onto the diagonals, \emph{i.e.}, evaluate the inner $\vd$ functions in \defref{def:vdb}.
	\end{minipage}
	\begin{minipage}{0.5\linewidth}
		\vspace{1ex}
        \resizebox{\linewidth}{!}{
\begin{tikzpicture}
    \matrix[
        matrix of math nodes,
        row sep=.8ex,
        column sep=1ex,
        left delimiter=(,right delimiter=),
        nodes={text width=.75em, text height=.5ex, text depth=.5em, align=center}
        ] (m) 
        {
        \alpha_1 & \gamma_2 & \beta_3 & \pi_4 & \kappa_5 & \omega_6 & \epsilon_7 & \eta_8 & \zeta_9\\
        \beta_1 & \alpha_2 & \gamma_3 & \omega_4 & \pi_5 & \kappa_6 & \zeta_7 & \epsilon_8 & \eta_9\\
        \gamma_1 & \beta_2 & \alpha_3 & \kappa_4 & \omega_5 & \pi_6 & \eta_7 & \zeta_8 & \epsilon_9\\
        \epsilon_1 & \eta_2 & \zeta_3 & \alpha_4 & \gamma_5 & \beta_6 & \pi_7 & \kappa_8 & \omega_9\\
        \zeta_1 & \epsilon_2 & \eta_3 & \beta_4 & \alpha_5 & \gamma_6 & \omega_7 & \pi_8 & \kappa_9\\
        \eta_1 & \zeta_2 & \epsilon_3 & \gamma_4 & \beta_5 & \alpha_6 & \kappa_7 & \omega_8 & \pi_9\\
        \pi_1 & \kappa_2 & \omega_3 & \epsilon_4 & \eta_5 & \zeta_6 & \alpha_7 & \gamma_8 & \beta_9\\
        \omega_1 & \pi_2 & \kappa_3 & \zeta_4 & \epsilon_5 & \eta_6 & \beta_7 & \alpha_8 & \gamma_9\\
        \kappa_1 & \omega_2 & \pi_3 & \eta_4 & \zeta_5 & \epsilon_6 & \gamma_7 & \beta_8 & \alpha_9\\
        };
        \draw[dashed, thin, black] (m-1-4.north west) -- (m-9-4.south west);
        \draw[dashed, thin, black] (m-1-7.north west) -- (m-9-7.south west);
        \draw[dashed, thin, black] (m-3-1.south west) -- (m-3-9.south east);
        \draw[dashed, thin, black] (m-6-1.south west) -- (m-6-9.south east);

        \draw[orange,rounded corners,rotate around={135:(-1.9,1.95)}] (-2.9,1.8) rectangle (-0.9,2.1);

        \draw[blue,rounded corners,rotate around={135:(-2.25,1.65)}] (-2.9,1.5) rectangle (-1.6,1.8);
        \draw[blue,rounded corners,rotate around={135:(-1.3,2.55)}] (-1.5,2.4) rectangle (-1.1,2.7);

        \draw[red,rounded corners,rotate around={135:(-2.6,1.35)}] (-2.8,1.2) rectangle (-2.4,1.5);
        \draw[red,rounded corners,rotate around={135:(-1.6,2.25)}] (-2.2,2.1) rectangle (-1,2.4);
        \begin{scope}[on background layer]
            \node[fit=(m-1-7)(m-3-9), fill=yellow!35, rounded corners] {};
            \node[fit=(m-1-4)(m-3-6), fill=cyan!25, rounded corners ] {};
            \node[fit=(m-1-1)(m-3-3), fill=green!15, rounded corners ] {};
            \node[fit=(m-4-7)(m-6-9), fill=cyan!35, rounded corners] {};
            \node[fit=(m-4-4)(m-6-6), fill=green!25, rounded corners] {};
            \node[fit=(m-4-1)(m-7-3), fill=yellow!15, rounded corners] {};
            \node[fit=(m-7-7)(m-9-9), fill=green!35, rounded corners ] {};
            \node[fit=(m-7-4)(m-9-6), fill=yellow!25, rounded corners  ] {};
            \node[fit=(m-7-1)(m-9-3), fill=cyan!15, rounded corners  ] {};
        \end{scope}

\end{tikzpicture}
}
	\end{minipage}%
	\hspace{0.5ex}
	\begin{minipage}{0.48\linewidth}
		Step 3: Regard the sub-arrays as entries of an $\mu \times \mu$ array and rearrange them on the diagonals, \emph{i.e.}, evaluate the outer $\vd$ function in \defref{def:vdb}.
	\end{minipage}
	\caption{Illustration of the modified vector-diagonal construction.}\label{fig:mvdeg}
\end{figure}
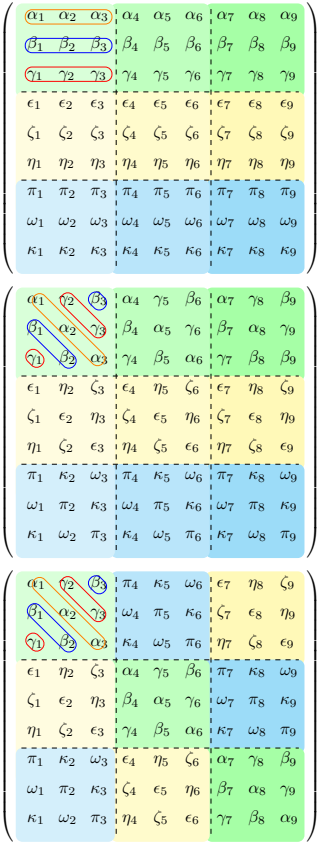

\defref{def:coverlocality} gives the straightforward generalization of locality in block codes to codes in the cover-metric, as the cover-metric codes consider lines as the units in which errors/erasures occur. We now give a more restrictive definition of locality in cover-metric, referred to as \emph{block-locality}, which can no longer be defined solely on lines. 
\begin{definition}[Block-Locality]\label{def:blocklocality}
Let $\Gamma^r$ and $\Gamma^c$ be partitions of $[n]$ with $|\Gamma^r_i| \leq r+\rho-1$ and $|\Gamma^c_i| \leq r+\rho-1$. An $[n\times n,k,d]_q^C$ array code $\mathbb{C}$ is said to have $(r,\rho)$ block-locality if for each symbol $c_{i,j},i\in[n],j\in[n]$ of a codeword $\mathbf{C}\in\mathbb{C}$, there exist two indices $a$ and $b$ such that
  \begin{enumerate}
    \item $i\in\Gamma^r_a$, $j\in\Gamma^c_b$
    \item $d_{C}(\mathbb{C}|_{\Gamma(a,b)}) \geq \rho$
  \end{enumerate}
 where
 \begin{equation*}
 \Gamma(a,b) = \{(s,l): s \in \Gamma^r_a, l\in \Gamma^c_b \} .
 \end{equation*}
\end{definition}
By this definition, the codeword arrays are divided into smaller local arrays of size at most $n_l\times n_l$ on which a cover distance $\rho$ is guaranteed. 
\begin{lemma}
A code $\mathbb{C}$ given by \defref{def:ArrayLRC} has $(r,\rho)$ block-locality as defined in \defref{def:blocklocality}.
\end{lemma}
\begin{proof}
The $\vdb$ function as in \defref{def:vdb} is a concatenation of an outer and $\mu^2$ inner $\vd$ functions. With the input as in \defref{def:ArrayLRC}, each inner $\vd$ returns a codeword of an $n_l \times n_l$ array code with cover distance at least $\rho$, as its inputs are codewords of a local code of an $[n,k',d,r,\rho]^H_q$ block LRC. As the outer $\vd$ function only rearranges the outputs of the inner $\vd$ functions, the code fulfills all requirements of block locality in \defref{def:blocklocality}. 
\end{proof}
In the following, we give an intuitive understanding of the construction in \defref{def:ArrayLRC} by virtue of an example.
\begin{example}[Modified vector-diagonal construction]
\label{eg:vdbprocess}
The modified vector-diagonal construction can be divided into three steps, which are shown in \figref{fig:mvdeg} with a $[9,4,5,2,2]_q^H$ LRC as constituent code. 
Such a code can be constructed e.g., by \cite{AFamily} when $q \geq n$.
In this example, $n_l=2+2-1=3$ and $\mu=\frac{9}{3}=3$. For simplicity we assume that the symbols in the same local group are in consecutive positions and indicate the local groups with different colors/shading. For readability we denote symbols by Greek letters, e.g. $\mathbf{c}=(\alpha_1,\hdots,\alpha_9)$. The array at Step 3 is a codeword of the  $[9\times 9, 36, 5, 2,2]_q^C$ cover-metric LRC.\\
\end{example}
\vspace{-15pt}
\section{Performance Comparison}\label{sec:prob}
In this section, we compare the decoding capabilities of our construction with the construction based on rank-metric codes~\cite{S.Kadhe} when considering a channel that corrupts lines of codewords. We say that $t$ (line-)errors occurred if the error matrix $\mathbf{E} \in \field^{n\times n}_q$ is of weight $\wt_C(\mathbf{E}) = t$. Similarly for erasures we say that $t$ (line-)erasures occurred if all erased positions can be covered by a cover $\mathcal{X}$ with $|\mathcal{X}| = t$. For ease of notation we define the \emph{erasure matrix} $\mathbf{E}$ to be any matrix in which all positions covered by $\mathcal{X}$ are non-zero.
We denote by $\mathbf{E}_{ij}$, with $i,j\in[\mu]$, the restriction of the matrix $\mathbf{E}$ to the local codes of size $n_l\times n_l$.\par
In the following we will compare the probability of an error/erasure pattern being locally decodable. We first consider erasures, the adaptation to errors is straight forward. If it holds that
\begin{equation}
\label{eq:conditionlocal}
\wt_C(\mathbf{E})\leq \rho-1,
\end{equation}
all erasures can be corrected within the local groups. For larger weights, the probability of local decoding being possible depends on which definition of locality the code fulfills. While our construction has locality in both, rows and columns, the construction of \cite{S.Kadhe} only has locality in the columns. It follows that for a rank-metric code with locality, any error/erasure in the rows necessarily intersects all local codes, as all of them span all rows.
\begin{example}\label{eg:erasurecomparison}
Consider an $[9\times 9,36,5,2,2]_q^C$ code with cover-locality as in \defref{def:coverlocality} (compare also Example~\ref{eg:vdbprocess}). This code guarantees to correct erasures of cover weight $d-1=4$ as the minimum cover distance of the code is $d=5$. The minimum cover distance of the local codes is $\rho=2$, so only ${\rho-1=1}$ line is guaranteed to be corrected locally. However, with high probability many error/erasure patterns of larger weight can still be corrected within the local codes, if the errors/erasures are distributed accordingly. \figref{fig:erasures} gives two examples of erasure patterns and illustrates the local decoding capability of the two code constructions.\\
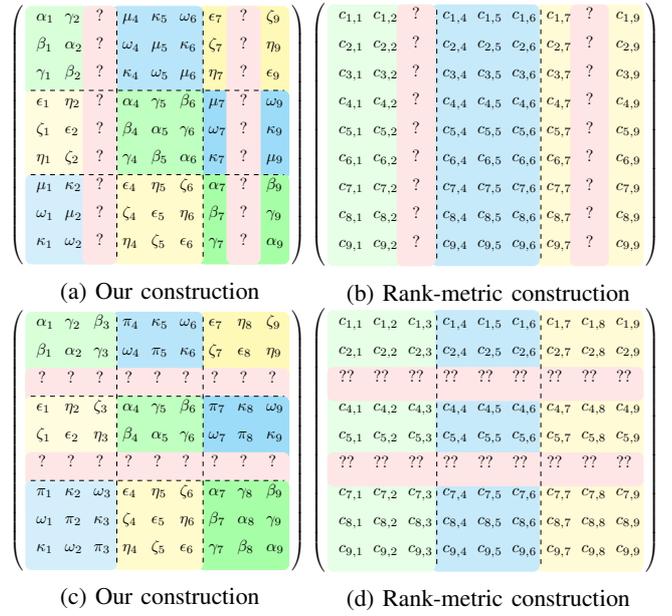
\begin{figure}
\centering
\begin{subfigure}[h]{0.46\linewidth}
      \centering
       \resizebox{\linewidth}{!}{
\begin{tikzpicture}
    \matrix(m)[
        matrix of math nodes,
        row sep=.8ex,
        column sep=.8ex,
        left delimiter=(,right delimiter=),
        nodes={text width=.75em, text height=.5ex, text depth=.5em, align=center}
        ]  
        {
        \alpha_1 & \gamma_2 & ? & \mu_4 & \kappa_5 & \omega_6 & \epsilon_7 & ? & \zeta_9\\
        \beta_1 & \alpha_2 & ? & \omega_4 & \mu_5 & \kappa_6 & \zeta_7 & ? & \eta_9\\
        \gamma_1 & \beta_2 & ? & \kappa_4 & \omega_5 & \mu_6 & \eta_7 & ? & \epsilon_9\\
        \epsilon_1 & \eta_2 & ? & \alpha_4 & \gamma_5 & \beta_6 & \mu_7 & ? & \omega_9\\
        \zeta_1 & \epsilon_2 & ? & \beta_4 & \alpha_5 & \gamma_6 & \omega_7& ? & \kappa_9\\
        \eta_1 & \zeta_2 & ? & \gamma_4 & \beta_5 & \alpha_6 & \kappa_7 & ? & \mu_9\\
        \mu_1 & \kappa_2 & ? & \epsilon_4 & \eta_5 & \zeta_6 & \alpha_7 & ? & \beta_9\\
        \omega_1 & \mu_2 & ? & \zeta_4 & \epsilon_5 & \eta_6 & \beta_7 & ? & \gamma_9\\
        \kappa_1 & \omega_2 & ? & \eta_4 & \zeta_5 & \epsilon_6 & \gamma_7 & ? & \alpha_9\\
        };
        \draw[dashed, thin, black] (m-1-4.north west) -- (m-9-4.south west);
        \draw[dashed, thin, black] (m-1-7.north west) -- (m-9-7.south west);
        \draw[dashed, thin, black] (m-3-1.south west) -- (m-3-9.south east);
        \draw[dashed, thin, black] (m-6-1.south west) -- (m-6-9.south east);
        \begin{scope}[on background layer]
            \node[fit=(m-1-7)(m-3-9), fill=yellow!35, rounded corners] {};
            \node[fit=(m-1-4)(m-3-6), fill=cyan!25, rounded corners ] {};
            \node[fit=(m-1-1)(m-3-3), fill=green!15, rounded corners ] {};
            \node[fit=(m-4-7)(m-6-9), fill=cyan!35, rounded corners] {};
            \node[fit=(m-4-4)(m-6-6), fill=green!25, rounded corners] {};
            \node[fit=(m-4-1)(m-7-3), fill=yellow!15, rounded corners] {};
            \node[fit=(m-7-7)(m-9-9), fill=green!35, rounded corners ] {};
            \node[fit=(m-7-4)(m-9-6), fill=yellow!25, rounded corners  ] {};
            \node[fit=(m-7-1)(m-9-3), fill=cyan!15, rounded corners  ] {};
            
            \node[fit=(m-1-3)(m-9-3), fill=red!10, rounded corners   ] {};
            \node[fit=(m-1-8)(m-9-8), fill=red!10 ,rounded corners  ] {};
        \end{scope} 
\end{tikzpicture}
}
      \caption{Our construction}
      \label{fig:egcolincmc}
\end{subfigure}%
\begin{subfigure}[h]{0.54\linewidth}
      \centering
       \resizebox{\linewidth}{!}{
\begin{tikzpicture}
    \matrix(m2)[
        matrix of math nodes,
        row sep=.8ex,
        column sep=1ex,
        left delimiter=(,right delimiter=),
        nodes={text width=1em, text height=.5ex, text depth=.5em, align=center}
        ]
        {
        c_{1,1} & c_{1,2} & ? & c_{1,4}  & c_{1,5} & c_{1,6} & c_{1,7} & ? & c_{1,9}\\
        c_{2,1} & c_{2,2} & ? & c_{2,4}  & c_{2,5} & c_{2,6} & c_{2,7} & ? & c_{2,9}\\
        c_{3,1} & c_{3,2} & ? & c_{3,4}  & c_{3,5} & c_{3,6} & c_{3,7} & ? & c_{3,9}\\
        c_{4,1} & c_{4,2} & ? & c_{4,4}  & c_{4,5} & c_{4,6} & c_{4,7} & ? & c_{4,9}\\
        c_{5,1} & c_{5,2} & ? & c_{5,4}  & c_{5,5} & c_{5,6} & c_{5,7} & ? & c_{5,9}\\
        c_{6,1} & c_{6,2} & ? & c_{6,4}  & c_{6,5} & c_{6,6} & c_{6,7} & ? & c_{6,9}\\
        c_{7,1} & c_{7,2} & ? & c_{7,4}  & c_{7,5} & c_{7,6} & c_{7,7} & ? & c_{7,9}\\
        c_{8,1} & c_{8,2} & ? & c_{8,4}  & c_{8,5} & c_{8,6} & c_{8,7} & ? & c_{8,9}\\
        c_{9,1} & c_{9,2} & ? & c_{9,4}  & c_{9,5} & c_{9,6} & c_{9,7} & ? & c_{9,9}\\
        };
        \draw[dashed, thin, black] (m2-1-4.north west) -- (m2-9-4.south west);
        \draw[dashed, thin, black] (m2-1-7.north west) -- (m2-9-7.south west);
        \begin{scope}[on background layer]
            \node[fit=(m2-1-1)(m2-9-3),  fill=green!10,rounded corners] {};
            \node[fit=(m2-1-7)(m2-9-9),  fill=yellow!20,rounded corners] {};
            \node[fit=(m2-1-4)(m2-9-6),  fill=cyan!20,rounded corners] {};
            
            \node[fit=(m2-1-3)(m2-9-3), fill=red!10,rounded corners   ] {};
            \node[fit=(m2-1-8)(m2-9-8), fill=red!10,rounded corners   ] {};
        \end{scope} 
\end{tikzpicture}
}
      \caption{Rank-metric construction}
      \label{fig:egcolinrmc}
\end{subfigure}
\begin{subfigure}[h]{0.46\linewidth}
	\centering
	   \resizebox{\linewidth}{!}{
    \begin{tikzpicture}
        \matrix(m)[
            matrix of math nodes,
            row sep=.8ex,
            column sep=.8ex,
            left delimiter=(,right delimiter=),
            nodes={text width=.75em, text height=.5ex, text depth=.5em, align=center}
            ]  
            {
            \alpha_1 & \gamma_2 & \beta_3 & \pi_4 & \kappa_5 & \omega_6 & \epsilon_7 & \eta_8 & \zeta_9\\
            \beta_1 & \alpha_2 & \gamma_3 & \omega_4 & \pi_5 & \kappa_6 & \zeta_7 & \epsilon_8 & \eta_9\\
            ?       &   ?       &   ?     & ?       &   ?     &     ?   &   ?       &   ?       &   ?  \\
            \epsilon_1 & \eta_2 & \zeta_3 & \alpha_4 & \gamma_5 & \beta_6 & \pi_7 & \kappa_8 & \omega_9\\
            \zeta_1 & \epsilon_2 & \eta_3 & \beta_4 & \alpha_5 & \gamma_6 & \omega_7 & \pi_8 & \kappa_9\\
            ?       &   ?       &   ?     & ?       &   ?     &     ?   &   ?       &   ?       &   ?  \\
            \pi_1 & \kappa_2 & \omega_3 & \epsilon_4 & \eta_5 & \zeta_6 & \alpha_7 & \gamma_8 & \beta_9\\
            \omega_1 & \pi_2 & \kappa_3 & \zeta_4 & \epsilon_5 & \eta_6 & \beta_7 & \alpha_8 & \gamma_9\\
            \kappa_1 & \omega_2 & \pi_3 & \eta_4 & \zeta_5 & \epsilon_6 & \gamma_7 & \beta_8 & \alpha_9\\
            };
            \draw[dashed, thin, black] (m-1-4.north west) -- (m-9-4.south west);
            \draw[dashed, thin, black] (m-1-7.north west) -- (m-9-7.south west);
            \draw[dashed, thin, black] (m-3-1.south west) -- (m-3-9.south east);
            \draw[dashed, thin, black] (m-6-1.south west) -- (m-6-9.south east);
            \begin{scope}[on background layer]
                \node[fit=(m-1-7)(m-3-9), fill=yellow!35,rounded corners] {};
                \node[fit=(m-1-4)(m-3-6), fill=cyan!25 ,rounded corners] {};
                \node[fit=(m-1-1)(m-3-3), fill=green!15 ,rounded corners] {};

                \node[fit=(m-3-1)(m-3-9), fill=red!10   ,rounded corners] {};

                \node[fit=(m-4-7)(m-6-9), fill=cyan!35,rounded corners] {};
                \node[fit=(m-4-4)(m-6-6), fill=green!25,rounded corners] {};
                \node[fit=(m-4-1)(m-7-3), fill=yellow!15,rounded corners] {};

                \node[fit=(m-6-1)(m-6-9), fill=red!10  ,rounded corners ] {};

                \node[fit=(m-7-7)(m-9-9), fill=green!35 ,rounded corners] {};
                \node[fit=(m-7-4)(m-9-6), fill=yellow!25 ,rounded corners ] {};
                \node[fit=(m-7-1)(m-9-3), fill=cyan!15 ,rounded corners ] {};
                
            \end{scope} 
    \end{tikzpicture}
}
      \caption{Our construction}
      \label{fig:egrowincmc}
\end{subfigure}%
\begin{subfigure}[h]{0.54\linewidth}
      \centering
       \resizebox{\linewidth}{!}{
    \begin{tikzpicture}
        \matrix(m2)[
            matrix of math nodes,
            row sep=.8ex,
            column sep=1ex,
            left delimiter=(,right delimiter=),
            nodes={text width=1em, text height=.5ex, text depth=.5em, align=center}
            ]
            {
            c_{1,1} & c_{1,2} & c_{1,3} & c_{1,4}  & c_{1,5} & c_{1,6} & c_{1,7} & c_{1,8} & c_{1,9}\\
            c_{2,1} & c_{2,2} & c_{2,3} & c_{2,4}  & c_{2,5} & c_{2,6} & c_{2,7} & c_{2,8} & c_{2,9}\\
                ??  &     ??  &     ??  &     ??   &     ??  &     ??  &     ??  &     ??  &     ?? \\
            c_{4,1} & c_{4,2} & c_{4,3} & c_{4,4}  & c_{4,5} & c_{4,6} & c_{4,7} & c_{4,8} & c_{4,9}\\
            c_{5,1} & c_{5,2} & c_{5,3} & c_{5,4}  & c_{5,5} & c_{5,6} & c_{5,7} & c_{5,8} & c_{5,9}\\
                ??  &     ??  &     ??  &     ??   &     ??  &     ??  &     ??  &     ??  &     ?? \\
            c_{7,1} & c_{7,2} & c_{7,3} & c_{7,4}  & c_{7,5} & c_{7,6} & c_{7,7} & c_{7,8} & c_{7,9}\\
            c_{8,1} & c_{8,2} & c_{8,3} & c_{8,4}  & c_{8,5} & c_{8,6} & c_{8,7} & c_{8,8} & c_{8,9}\\
            c_{9,1} & c_{9,2} & c_{9,3} & c_{9,4}  & c_{9,5} & c_{9,6} & c_{9,7} & c_{9,8} & c_{9,9}\\
            };
            \draw[dashed, thin, black] (m2-1-4.north west) -- (m2-9-4.south west);
	        \draw[dashed, thin, black] (m2-1-7.north west) -- (m2-9-7.south west);
            \begin{scope}[on background layer]
                
                \node[fit=(m2-1-7)(m2-9-9),  fill=yellow!20,rounded corners] {};
                \node[fit=(m2-1-4)(m2-9-6),  fill=cyan!20,rounded corners] {};
                \node[fit=(m2-1-1)(m2-9-3),  fill=green!10,rounded corners] {};
                \node[fit=(m2-3-1)(m2-3-9), fill=red!10  ,rounded corners ] {};
                \node[fit=(m2-6-1)(m2-6-9), fill=red!10   ,rounded corners] {};
            \end{scope} 
    \end{tikzpicture}

}
      \caption{Rank-metric construction}
      \label{fig:egrowinrmc}
\end{subfigure}
\caption{Illustration of erasure patterns. Locally repairable erasures are denoted by `?' while globally repairable but not locally repairable erasures are denoted by `??'.}\label{fig:erasures}
\end{figure}

\end{example}
\vspace{-15pt}
\subsection{Probability of Local Decoding}
To analyze the probability of local decoding being possible we define two recursive functions that count the number of crisscross error/erasure patterns that can be locally corrected. 
\begin{definition}\label{def:S}
	Define the function
	\begin{equation*}
    \begin{aligned}
    	&S(t,t_l,g,a)\defeq\\
		&\left\{
		\begin{array}{ll}
		1, &t=0\\
		0, &t_l = 0,t>0 \\
		\begin{aligned}
		&\sum\limits_{i = \max\{a, t-g(t_l-1)\}}^{\min\{g,\floor{\frac{t}{t_l}}\}}\Bigg(\binom{g}{i}\binom{n_l}{t_l}^i\\
		&\quad S(t-it_l, t_l-1,g-i,0)\Bigg)
		\end{aligned},& \textrm{else}
		\end{array}
		\right.
    \end{aligned}
	\end{equation*}
\end{definition}
The function given in \defref{def:S} counts the error vectors of weight $t$ that can be locally corrected by the block LRCs with $g$ local groups where each group can correct at most $t_l$ corrupted symbols. The parameter $a$ is the minimum number of groups where $t_l$ symbols are corrupted.
\begin{definition}\label{def:Sarray}
Define the function
\begin{equation*}
\begin{aligned}
	&S_{\textrm{array}}(t,l,t_r,l_r,g)\defeq\\
    &\left\{
			\begin{array}{ll}
				S(t,l,g,0), &t_r = 0\\
				& \\
				\begin{aligned}
				&\binom{g}{1}\binom{n_l}{t_r}S(t-t_r,l-t_r,g,0)\\
				& + S_{array}(t,l,t_r,t_r-1,g) 
				\end{aligned},
				&0 < t_r\leq l_r \\
				& \\
				\begin{aligned}
				&S(t_r, l_r, g, n_l,1)S(t-t_r, l-l_r, g,0)\\
				&+ S_{array}(t,l,t_r,l_r-1,g)
				\end{aligned},
				&l_r < t_r\leq g l_r \\
				& \\
				0 ,&\textrm{else} \\
			\end{array}
		\right.
\end{aligned}
\end{equation*}
\end{definition}
The function given in \defref{def:Sarray} counts the error/erasure matrices of cover-weight $t$, where $t_r$ out of $t$ erroneous lines occur in rows, that can be locally corrected by a cover-metric LRC under \defref{def:coverlocality}, which has $2g$ local codes, \emph{i.e.}, $g$ local codes consisting of rows and $g$ consisting of columns. The parameter $l$ is the decoding capability of local codes when $l_r$ erroneous lines are rows.\\
For any cover-metric code $[n\times n,k, d,r,\rho]_q^C$ as defined in \defref{def:ArrayLRC}, the probability $p_l(t)$ of locally decoding crisscross errors of cover-weight $t$ is given by
\vspace{-5pt}
\begin{equation*}
p_{l}(t) = \frac{\sum\limits_{t_r = 0}^{t}S_{\textrm{array}}(t, \floor{\frac{\rho-1}{2}}, t_r, \floor{\frac{\rho-1}{2}}, \mu)}{\binom{2n}{t}}. 
\end{equation*}
For the code construction based on rank-metric codes~\cite{S.Kadhe} the probability is given by
\begin{equation*}
p_{l}(t) = \frac{\sum\limits_{t_r = 0}^{\left\lfloor \frac{\rho-1}{2}\right\rfloor}\binom{n}{t_r}S(t-t_r, \floor{\frac{\rho-1}{2}}-t_r, \mu,0)}{ \binom{2n}{t}}.
\end{equation*}
For the rank-metric construction, the distribution of erroneous rows does not affect the local decoding capability as long as $t_r\leq  \floor{\frac{\rho-1}{2}}$. Thus, after arbitrarily distributing $t_r$ into $n$ rows, we only need to consider the locally correctable column distributions, which can be counted by the $S(\cdot)$ function.

Notice that these two formula can be also used to calculate the probability of local decoding crisscross erasures by substituting $\floor{\frac{\rho-1}{2}}$ with $\rho-1$. \par
\begin{figure}
\hspace{-2em}
	\begin{subfigure}[h]{0.5\columnwidth}
        \resizebox{1.1\linewidth}{!}{
\begin{tikzpicture}
    \begin{axis}[
    title ={$[255\times 255,112\cdot 255, 53,8,8]$ ($n_l=15,\mu=17$)},
        grid=both,
        grid style={line width=.1pt, draw=gray!20},
        major grid style={line width=.2pt,draw=gray!50},
        xlabel={$t$},
        ylabel={$p_l(t)/\%$},
        minor x tick num=5,
        minor y tick num=5,
        legend entries={modified vd., rank-metric [13]},
        enlargelimits=false,
        legend pos=north east
        ]
        \addplot table [x index=0, y index=1] {data255.dat} ;
        \addplot table [x index=0, y index=2] {data255.dat} ;
        \draw[thick,dashed] (26,0) -- (26,100);
        \node at (22,60) {$t=\lfloor{\frac{d-1}{2}}\rfloor$};
    \end{axis}
\end{tikzpicture}

        }
	    \vspace{-20pt}
        \caption{\centering Locally decoding crisscross errors}
	\end{subfigure}\hspace{5pt}%
	\begin{subfigure}[h]{0.5\columnwidth}		
        \resizebox{1.1\linewidth}{!}{
\begin{tikzpicture}
    \begin{axis}[
    title ={$[255\times 255,112\cdot 255, 53,8,8]$ ($n_l=15,\mu=17$)},
        grid=both,
        grid style={line width=.1pt, draw=gray!20},
        major grid style={line width=.2pt,draw=gray!50},
        xlabel={$t$},
        ylabel={$p_l(t)/\%$},
        minor x tick num=10,
        minor y tick num=5,
        legend entries={modified vd., rank-metric [13]}, 
        enlargelimits=false,
        legend pos=north east
        ]
        \addplot table [x index=0, y index=1] {era255cover.dat} ;
        \addplot table [x index=0, y index=1] {era255rank.dat} ;
        \draw[thick,dashed] (52,0) -- (52,100);
        \node at (65,60) {$t=d-1$};
    \end{axis}
\end{tikzpicture}

        }
		  \vspace{-20pt}
	      \caption{\centering Locally decoding crisscross erasures}
	      \label{fig:resultera}
	\end{subfigure}
\vspace{-1ex}
\caption{Probability of locally decoding crisscross errors} \label{fig:prob_error_result}
\vspace{-15pt}
\end{figure}
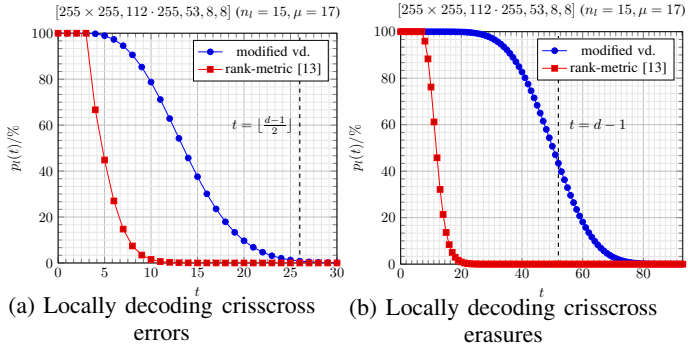
The probabilities for the code parameters $n = 255$, $k = 112$, $r = 8$ and $\rho = 8$ are shown in \figref{fig:prob_error_result}. It is apparent that our construction has higher probability of locally decoding crisscross errors/erasures than the construction based on rank-metric codes.

\section{Conclusion and Outlook}
In this work, we have generalized the notion of locality to the cover metric, proved an alphabet-free upper bound on the distance of a cover-metric code with locality and introduced a bound-achieving construction. Further, we compared the performance in terms of the probability of successful local decoding with the construction given in \cite{S.Kadhe}. 

We only considered bounded minimum distance decoding of the constituent codewords. It was shown in \cite{Wachter-Zeh-Crisscross} that codes in the cover metric can be list decoded up to a Johnson-like upper bound. In \cite{LHListLRC} it was shown that for some parameters block LRCs can be list decoded beyond the Johnson radius. Future work of interest includes the combination of these results as well as the use of list or interleaved decoders instead of a BMD decoder for the codewords of the constituent code.









\bibliographystyle{ieeetr}
\bibliography{cover_bib}

\end{document}